\documentclass[iicol,pdflatex,sn-mathphys-num]{sn-jnl}
\usepackage{graphicx}%
\usepackage{multirow}%
\usepackage{amsmath,amssymb,amsfonts}%
\usepackage{amsthm}%
\usepackage{mathrsfs}%
\usepackage[title]{appendix}%
\usepackage{xcolor}%
\usepackage{textcomp}%
\usepackage{manyfoot}%
\usepackage{booktabs}%
\usepackage{algorithm}%
\usepackage{algorithmicx}%
\usepackage{algpseudocode}%
\usepackage{listings}%
\usepackage[version=4]{mhchem}%
\usepackage{lineno}
\usepackage{ulem}
\usepackage{cancel}
\usepackage{soul}

\theoremstyle{thmstyleone}%

\theoremstyle{thmstyletwo}%
\theoremstyle{thmstylethree}%

\title{Non-Equilibrium Stratification in Supercritical \ce{CO2}}

\author*[1,2]{\fnm{Paul} \sur{Fruton}}\email{paul.fruton@unimi.it}
\author[1,3,4]{\fnm{Emma} \sur{Lisoir}}
\author[1,5]{\fnm{Happiness} \sur{Imuetinyan}}
\author[1,6]{\fnm{Cédric} \sur{Giraudet}}
\author[1]{\fnm{Fabrizio} \sur{Croccolo}}

\affil[1]{\orgname{Universite de Pau et des Pays de l'Adour, E2S UPPA, CNRS, LFCR}, \orgaddress{\city{Anglet}, \postcode{64600}, \country{France}}}
\affil[2]{\orgdiv{Dipartimento di Fisica 'A. Pontremoli'}, \orgname{Università degli Studi di Milano}, \orgaddress{\city{Milano}, \postcode{20133}, \country{Italy}}}
\affil[3]{\orgname{Univ Toulouse, INSA Toulouse, CNRS, LPCNO}, \orgaddress{\city{Toulouse}, \country{France}}}
\affil[4]{\orgname{Laboratoire de Génie Chimique, CNRS, Toulouse INP, Université de Toulouse}, \orgaddress{\city{Toulouse}, \country{France}}}
\affil[5]{\orgname{Universite de Pau et des Pays de l'Adour, E2S UPPA, CNRS, LFCR}, \orgaddress{\city{Pau}, \postcode{64000}, \country{France}}}
\affil[6]{\orgname{CTS Consulting \& Technical Support}, \orgaddress{\city{Vernon}, \postcode{27200}, \country{France}}}

\begin{document}

\abstract{Supercritical matter is commonly described as a continuous phase lacking sharp boundaries between liquid-like and gas-like regions. Under non-equilibrium conditions, however, this picture may fail. Here, we experimentally investigate non-equilibrium density fluctuations in supercritical carbon dioxide (\ce{CO2}) subjected to a stabilizing temperature gradient. Using shadowgraphy, we observe the spontaneous emergence of fluid stratification into distinct layers separated by transition regions, across which thermodynamic properties vary sharply. These signatures become particularly pronounced as the system crosses the Widom lines, where thermodynamic response functions exhibit extrema in the supercritical regime. Analysis of the structure function of fluctuations reveals Brunt–Väisälä oscillations at multiple frequencies, revealing gravity-driven coupling between thermal and viscous modes and direct evidence of hydrodynamic stratification. Our approach enables systematic exploration of a broad range of thermodynamic states within a single experiment. Altogether, these findings indicate that far from remaining homogeneous, the Widom region develops structured dynamical heterogeneity under non-equilibrium conditions.
}


\maketitle

\section*{Introduction}
Commonly, the supercritical phase is referred to as a phase that is neither gas nor liquid, where the fluid properties vary smoothly from those of a gas to those of a liquid. This combination allows a supercritical fluid to exhibit properties of both phases, such as the high density of the liquid phase and the low viscosity of the gas one, making the supercritical state of great interest for many industrial applications. A suitable example is the transport of supercritical \ce{CO2} through pipelines, where its high density allows efficient fluid transfer, while its low viscosity minimises energy consumption. 

This theoretical description holds under equilibrium conditions, but fluids in the real world are only exceptionally at equilibrium, because almost every real condition features inhomogeneities in the thermodynamic variables, leading the fluid out of equilibrium.
An additional question is: can a fluid out of equilibrium be used to test different points of the phase diagram as the superposition of distinct equilibrium points?

\begin{figure*} [h!]
    \centering
    \includegraphics[width=0.85\textwidth]{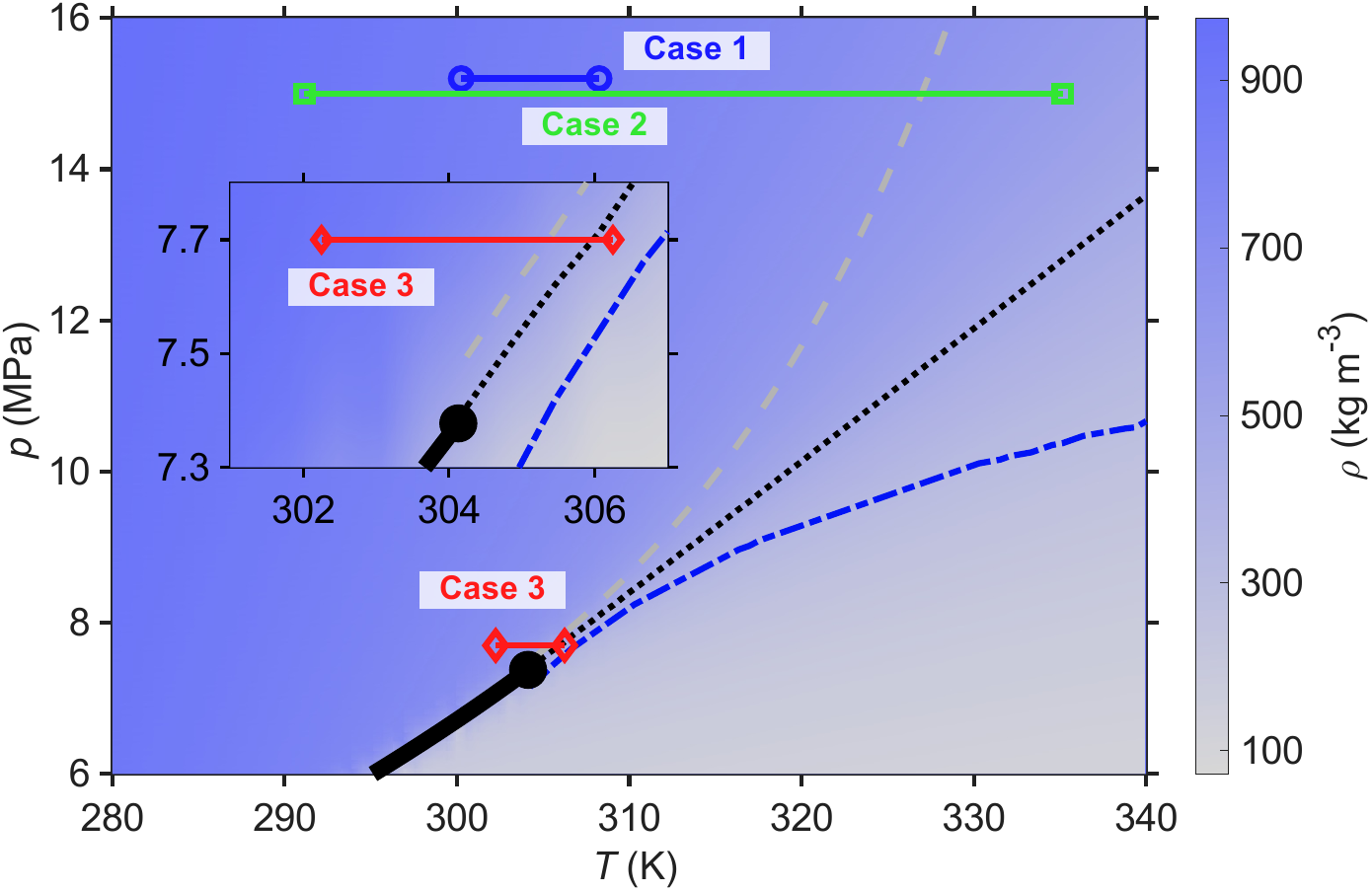}
    \caption{\textbf{Density map of pure \ce{CO2} in the $p$-$T$ diagram.} Indications of the Widom region limits are provided by the minimum sound speed line (dashed and dotted blue line) and the isenthalpy line (dashed gray line), while the isochore is displayed as well (dashed dark line). They emerge from the critical point (dark plain circle) ending the boiling line (dark plain line). The thermal gradients for the experiments analysed and discussed here are shown as solid horizontal segments: case 1 in blue ($p$ = 15 MPa, $T_{\mathrm{mean}}$ = 31.1 °C, $\Delta T$ = 8 K), case 2 in red ($p$ = 15 MPa, $T_{\mathrm{mean}}$ = 40 °C, $\Delta T$ = 44 K), and case 3 in red ($p$ = 7.7 MPa, $T_{\mathrm{mean}}$ = 31.1 °C, $\Delta T$ = 4 K). Inset: close-up of the $p$-$T$ diagram around the critical point.}
    \label{fig:densMap}
\end{figure*}

Regarding the description of the supercritical state, the concept of Widom lines has emerged in recent decades as a way to describe the loci of maxima in response functions, such as heat capacity, compressibility, or sound speed, extending into the supercritical region from the critical point \cite{fomin_thermodynamic_2015, mareev_optical_2020}. In this work, we use the term “Widom region” in an operational sense, referring to the near-critical domain where thermophysical properties exhibit strong variations at fixed pressure. In Fig.\ref{fig:densMap}, the loci of the extrema of response functions, retrieved or computed from NIST data, are all embodied by the isenthalpy and minimum sound speed lines, which thus draw the limits of the Widom region.

This notion should be distinguished from the Frenkel line, which describes a dynamical crossover between liquid-like and gas-like microscopic dynamics and persists far from criticality \cite{brazhkin_two_2012, bryk2017}. Our experiments probe near-critical supercritical \ce{CO2} under stabilizing temperature gradients, where the observed dynamics arise from buoyancy, thermophysical properties, and fluctuating hydrodynamics, rather than from a Frenkel-line crossover. This is also related to the fact that our technique investigates the mesoscopic scales in the frequency domain up to hundreds Hz, while the microscopic motion related to Frenkel-line motion would be at frequencies larger by many orders of magnitude. The multiple dynamical contributions reported here are therefore interpreted as hydrodynamic signatures of depth-dependent thermophysical properties, not as evidence of a Frenkel-line transition. The significance of the Widom lines, however, has rarely been tested under non-equilibrium conditions. Moreover, the equilibrium perspective neglects the role of fluctuations, which are ubiquitous in fluid systems, even under equilibrium conditions, and become increasingly important in the vicinity of the critical point or in the presence of a macroscopic gradient \cite{Ortiz2006, Croccolo2016}. 

Indeed, in a pure fluid stressed by a thermal gradient, giant nonequilibrium fluctuations (NEFs) take place at all possible spatial scales, or equivalently, wave numbers. NEFs are largely amplified by the coupling between vertical velocity fluctuations and the macroscopic gradient and are limited only by gravity or the finite size of the sample \cite{Vailati1996,Vailati1998,ortizdezarate_2020,Giraudet2015}.  
In this study, we focus on the intermediate scattering function (ISF) of temperature fluctuations, which characterizes their dynamics and the corresponding time evolution, providing an efficient insight into the processes that take place within the fluid \cite{croccolo_2006, Croccolo2012}.

Let us consider a thin fluid layer placed out of equilibrium by a stabilizing thermal gradient far from the critical point. In such conditions, the vertical density profile of the fluid is almost linear and the thermophysical properties show a small dependence with the height, so that only small variations occur around a base state ($T_0$, $p_0$). The response of thermodynamic variables to small fluctuations is therefore approximately linear and the system can be described within the framework of linearized fluctuating hydrodynamics (FHD). From the fluctuating Oberbeck-Boussinesq equations, we have access to the dynamic structure factor of density fluctuations \cite{Ortiz2006}, $S_\rho(q,\omega)$, which shares the same dynamic as the temperature fluctuating field as they are locally related by $\delta\rho/\rho_0\sim -\beta_T\delta T$. $S_\rho(q,\omega)$ is the temporal Fourier transform of the local time correlation function associated with density fluctuations at depth $z$ 
\begin{equation}
    C_z^{(\rho)}(q,dt)=\langle \delta \rho(q,z,dt)\,\delta \rho^*(q,z,0)\rangle,
\end{equation}
where $q$ is the wavenumber and $dt$ the time delay. 

Therefore, far from the critical point, the correlation function exhibits a single dominant decaying mode with a decay time $\tau(q)$ that depends both on thermal diffusivity and gravity \cite{schmitz1985,law1988,segre1992,Ortiz2006,Croccolo2016}

\begin{equation}
    \tau(q)=\frac{1}{\alpha_T q^2 \left[ 1+\left(\frac{q_{ro,T}}{q}\right)^4\right]},
\label{eq:tau_grav}
\end{equation}
where $q_{ro,T}$ is the characteristic roll-off wave number at which diffusion of temperature and gravity act on similar timescales \cite{croccolo_2006, Croccolo2012}

\begin{equation}
    q_{ro,T}=\left|\frac{\beta_T \ \boldsymbol{\mathrm{g}} \cdot \boldsymbol{\nabla} T}{\alpha_T \ \nu}\right|^{1/4}=\frac{|\mathrm{Ra}|^{1/4}}{L},
\end{equation}
with $\beta_T=-(1/\rho)\ (\partial\rho/\partial T)_p$ the isobaric thermal expansion coefficient, $\boldsymbol{\mathrm{g}}$ the acceleration of gravity, $\boldsymbol{\mathrm{\nabla}} T$ the temperature gradient across the sample, $\alpha_T$ the thermal diffusivity, $\nu$ the kinematic viscosity, $\mathrm{Ra}$ the Rayleigh number and $L$ the thickness of the fluid layer.

Large fluctuations ($q\ll q_{ro,T}$) relax faster by buoyancy, while small ones ($q\gg q_{ro,T}$) vanish faster by thermal conduction and $\tau(q)\simeq1/\alpha_Tq^2$ \cite{Ortiz2006,Croccolo2016}.
At small wave numbers, inertia leads to the coupling between thermal and viscous modes resulting in a propagative contribution predicted by fluctuating hydrodynamic theory \cite{Ortiz2006, Croccolo_2019}. For $q<q_p=\frac{1}{L}(-\frac{4\mathrm{Ra~Pr}}{(\mathrm{Pr}-1)^2})^{1/4}$, with $\mathrm{Pr}={\nu}/{\alpha_T}$ the Prandtl number, the time correlation function shows damped oscillations due to internal gravity waves, modelled by 
\begin{equation}
    C_z^{(\rho)}(q,t)\propto \mathrm{cos}(\Omega(q)~ t+\phi)~ \mathrm{exp}\left(-\frac{t}{\tau(q)}\right)
    \label{eq:sf_with_1_oscillation}
\end{equation}
where $\tau(q)\simeq 2/[(\nu + \alpha_T)q^2]$ is the decay time of the coupled mode and $\Omega(q)$ is the oscillation frequency bounded by the Brunt-Väisälä frequency \cite{Carpineti_2024, Ortiz2006,Croccolo_2019}
\begin{equation}
\Omega_{\mathrm{max}} = \sqrt{\boldsymbol{\mathrm{g}} \cdot \boldsymbol{\nabla}\rho/{\rho}} = \sqrt{\beta_T\ \boldsymbol{\mathrm{g}} \cdot \boldsymbol{\nabla} T}.
    \label{eq:BV_freq}
\end{equation}
Long-range correlated fluctuations oscillate within the sample, while those correlated over smaller distances decay either by buoyancy or diffusion without oscillating, as already observed in liquid mixtures of two or three components for temperature and concentration fluctuations \cite{Croccolo_2019, garciafernandez_2019, ortizdezarate_2020}.

In our experiments, we use a high-sensitivity shadowgraph setup to measure density non-equilibrium fluctuations in supercritical \ce{CO2} layers subjected to vertical stabilising temperature gradients. Shadowgraphy provides a depth-integrated fluctuating intensity directly related to the density, and thus, temperature, fluctuating fields
\begin{equation}
    \delta I(q,dt) = \int_0^LW(q,z)\delta\rho(q,z,dt)dz
\end{equation}
with $W(q,z)$ an optical transfer. This method has previously been applied to critical fluids and polymer solutions to study giant fluctuations, thermophysical properties, and convection under gravity \cite{Croccolo2016, Castellini_2024}. Challenging measurements have also been performed under microgravity conditions due to the robustness of the dynamic shadowgraph approach \cite{Vailati2011}.

\begin{figure*} [ht!]
\centering
    \includegraphics[width=0.95\textwidth]{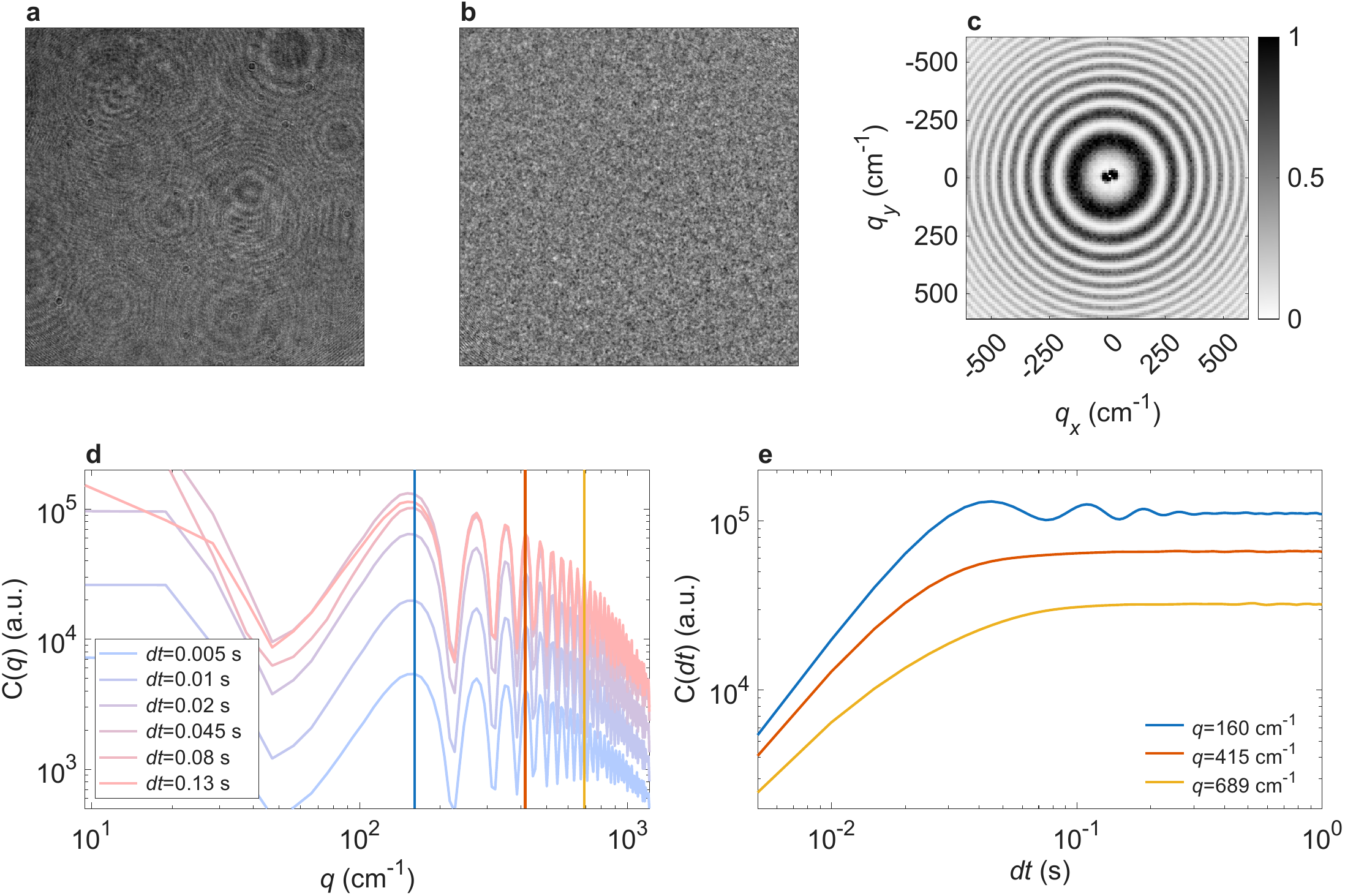}
    \caption{\textbf{Differential Dynamic Algorithm analysis of shadowgraph images.} Sample data  at $p$ = 7.5 MPa, $T_{\mathrm{mean}}$ = 31.1 °C, $\Delta T$ = 4 K. (a) Shadowgraph raw image, (b) image difference between two images acquired with a time delay $dt=5$ s, (c) normalised 2D-structure function of the density fluctuations for $dt=5$ s, normalised 1D-structure functions (d) for different $dt$ from 0.005 (light blue) to 0.13 (light red) s, and (e) for three different wave numbers $q$ of 160 (blue), 415 (orange), and 689 (yellow) $\mathrm{cm^{-1}}$.}
    \label{fig:images}
\end{figure*}

Using our methodology \cite{Croccolo2012}, we analyse density 2D-maps in the Fourier space, thus separating the signal at different wave numbers $q$.
By means of the Differential Dynamic Algorithm \cite{croccolo_2006, Croccolo2012}, whose concept is illustrated in Fig. \ref{fig:images}, we compute the structure functions (SFs) of fluid density fluctuations, captured by shadowgraphy, integrated over the vertical axis, parallel to the temperature gradient. The SFs contain quantitative information on the evolution of NEFs in both time and space, according to the following equation

\begin{equation}
    \mathrm{SF}(q,dt)=2\{S(q)~ T(q) [1-f(q,dt)]+B(q)\},
\end{equation}
where $S(q)$ is the NEF static structure function, $T(q)$ is the shadowgraph transfer function, $f(q,dt)$ is the ISF and $B(q)$ is a noise term \cite{Castellini_2024}. The ISF is defined as the normalized autocorrelation function of the fluctuating intensity field $\delta I(q,dt)$. Using the fact that the cross-correlations $\langle\delta\rho(q,z,t)\delta\rho^*(q,z',0)\rangle$ decay rapidly with $|z-z'|$ in the stably stratified conduction state, we obtain
\begin{equation}
    f(q,dt) \simeq \frac{\int_0^LW^2(q,z)C_z^{(\rho)}(q,dt)dz}{\int_0^LW^2(q,z)C_z^{(\rho)}(q,0)dz}.
    \label{eq:ISFint}
\end{equation}
When the thermophysical properties vary weakly with depth, $C_z^{(\rho)}(q,dt)$ can be approximated as independent of $z$, yielding a single effective contribution described by Equation (\ref{eq:sf_with_1_oscillation}). This corresponds to the usual situation of a thin fluid layer under a stabilizing thermal gradient, where the temperature, density, and thermophysical coefficients vary continuously across the sample but weakly enough so that the system can be treated as quasi-homogeneous around a single base state $(T_0,p_0)$, typically taken at the mean temperature.

In contrast, when thermophysical properties vary strongly with height, the quasi-homogeneous approximation is no longer valid, but can be still considered locally valid in thinner sublayers. Within the experimental resolution, we approximate this continuously stratified system by a finite number $N$ of effective sublayers $\mathcal{R}_i$, where the local thermophysical properties can be treated as constant. The ISF can be written as
\begin{equation}
    f(q,dt) = \sum_{i=1}^N A_i(q)\frac{\mathrm{cos}(\Omega_i dt+\phi_i)}{\mathrm{cos}(\phi_i)}\mathrm{exp}\left(-\frac{dt}{\tau_i(q)}\right),
    \label{eq:ISFdisc}
\end{equation}
where $\Omega_i(q)$ and $\tau_i(q)$ are the oscillation frequency and decay time associated with the effective sublayer $\mathcal{R}_i$, and $A_i(q)\propto \int_{\mathcal{R}_i}W^2(q,z)C_z^{(\rho)}(q,0)\,dz$ is the weight of region $\mathcal{R}_i$, normalized such that $\sum_i A_i(q)=1$.

In practice, $N$ represents the minimum number of quasi-homogeneous sublayers required to approximate the depth-dependent profiles of $\beta_T(z)$, $\alpha_T(z)$, and $\nu(z)$ within the experimental resolution. Far from the Widom region, a single effective layer is sufficient ($N=1$), whereas close to or inside the Widom region, the stronger depth dependence of the thermophysical properties requires $N>1$ effective sublayers. These contributions should be understood as a reduced, piecewise quasi-homogeneous representation of a continuously stratified system. They do not imply the existence of discrete equilibrium thermodynamic phases, but rather reflect the depth-dependent hydrodynamic response of the system.

In our experiments, we use a specialised high-pressure cell (HP) to investigate temperature fluctuations in a layer of pure \ce{CO2} stressed by a stabilising temperature gradient. 
Details on the HP cell and optical setup can be found in the Methods section and in previous publications \cite{giraudet_high-pressure_2014,fruton2023}.

Here we report the results of three significant experiments with different average temperatures, pressures, and temperature gradients, as shown in Fig. \ref{fig:densMap}, where the limits of the Widom region, defined by the minimum speed-of-sound line and the isenthalpy line, are also reported.
The colour-code also shows the different values of fluid density, calculated from NIST data \cite{Span1996}.

In regions where equilibrium response functions vary sharply with temperature at fixed pressure, as in the Widom region, the coefficients entering the linearized Oberbeck--Boussinesq description, notably $\beta_T$, $\alpha_T$, and $\nu$, are strongly depth-dependent. Because the Brunt--Väisälä frequency and the fluctuation decay times depend explicitly on these coefficients (Equations (\ref{eq:BV_freq}) and (\ref{eq:tau_grav})), the non-equilibrium steady state is expected to exhibit a stratified dynamical response. The number of effective contributions retained in each case is selected with the robust fitting procedure described in the Methods section.

\section*{Results}
\bmhead{Case 1}

\begin{figure*} [ht!]
\centering
    \includegraphics[width=0.99\textwidth]{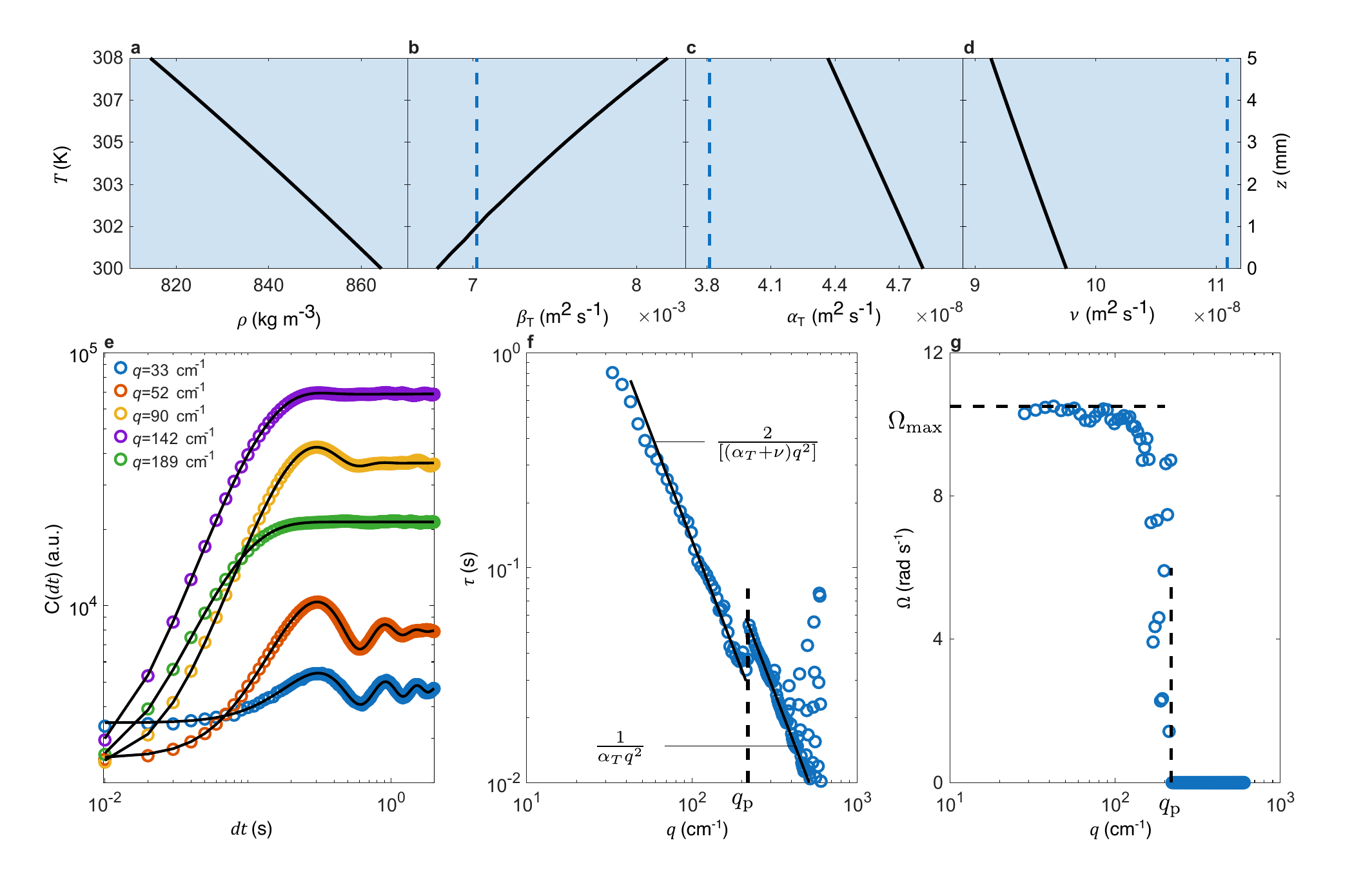}
    \caption{\textbf{Experimental results for case 1 under thermodynamic conditions far from the Widom region.} Data at $p$ = 15 MPa, $T_{\mathrm{mean}}$ = 31.1 °C, $\Delta T$ = 8 K, blue gradient in Fig.\ref{fig:densMap}. 
    (a-d) Thermophysical properties of the system: (a) density, (b) thermal expansion coefficient, (c) thermal diffusivity, and (d) kinematic viscosity. All the properties are plotted against temperature or height, assuming a linear relationship between them. The temperature and height are on the vertical axis for all graphs. The vertical dotted lines show the measured data from the fitting of the SFs, as explained in the text.
    (e) 1D-Structure functions for 5 wave numbers (33 (blue), 52 (orange), 90 (yellow), 142 (purple), and 189 (green) $\mathrm{cm^{-1}}$) and their fits (dark line). (f) Time decay $\tau(q)$ of the density fluctuations and (g) Brunt-Väisälä frequency $\Omega(q)$ as obtained by fitting the structure functions shown in (e) with the model of Equation (\ref{eq:sf_with_1_oscillation}).}
    \label{fig:SF1}
\end{figure*}

The first experiment is carried out far from the Widom region ($p$ = 15 MPa, $T_{\mathrm{mean}}$ = 31.1 °C, $\Delta T$ = 8 K, blue thermal gradient in Fig.\ref{fig:densMap}). Under these conditions, the density evolves quasi-linearly and the thermal expansion coefficient, the thermal diffusivity, and the dynamic viscosity are almost constant:  $\Delta_{\mathrm{rel}}\beta_T = 19\%$, $\Delta_{\mathrm{rel}}\alpha_T =10\%$, $\Delta_{\mathrm{rel}}\nu =7\%$, cf. Fig. \ref{fig:SF1} (a), with $\Delta_{\mathrm{rel}}X=(X_{\mathrm{max}}-X_{\mathrm{min}})/\bar{X}$ and $\bar{X}$ the mean value. The resulting SFs are plotted in Fig. \ref{fig:SF1} (e). At $q<q_p$, they contain one thermal mode whose amplitude is modulated by a sinusoidal function, which is the signature of a propagative mode \cite{Croccolo_2019, garciafernandez_2019, ortizdezarate_2020} as described by Equation (\ref{eq:sf_with_1_oscillation}) and as already reported for thermal and solutal fluctuations in liquid mixtures \cite{Croccolo_2019,garciafernandez_2019}. The propagative mode results from the coupling between the thermal and viscous modes. The latter is not directly measurable by optical means, as it does not contribute to density fluctuations in the system, but gravity couples it to the thermal mode, producing oscillations at the mesoscopic scale within the sample for wave numbers $q<q_p$. The apparent discontinuity in $\tau(q)$ at $q \simeq q_p$ reflects a crossover between coupled and decoupled hydrodynamic regimes and does not correspond to a physical singularity.

By fitting the structure functions over the entire range of $q$ through the model provided by Equation (\ref{eq:sf_with_1_oscillation}), we obtain both the time decay $\tau(q)$ and the frequency $\Omega(q)$, as reported in panels (f) and (g) of Fig. \ref{fig:SF1}, respectively. For large wave number $q>q_p$, the time decay expressed by Equation (\ref{eq:tau_grav}) can be approximated by the classical thermal diffusive equation  $\tau(q)=1/(\alpha_T q^2)$ to obtain a measurement of the fluid thermal diffusivity $\alpha_T$. 
The resulting fit is shown in panel (f) as a dark line, corresponding to a thermal diffusivity $\alpha_T=(3.9 \pm 0.1)\times 10^{-8}~\mathrm{m^2~ s^{-1}}$, a value reported with the vertical dashed blue line in Fig. \ref{fig:SF1}(c).
For wave numbers $q<q_p$, the model for the time decay includes the kinematic viscosity, due to the coupling between the thermal mode and the viscous one. The time decay in this case can be modelled by \cite{Wu2020} $\tau(q)=2/[(\alpha_T+\nu) q^2]$ which provides an estimate of the fluid kinematic viscosity $\nu=(11.1 \pm 0.5)\times 10^{-8}~\mathrm{m^2~ s^{-1}}$, as shown with the dashed blue line in panel (d).

Additionally, the fit of the SFs provides the values of the oscillating frequency for $q<q_p$, as shown in panel (g), from which one can estimate the maximum Brunt-Vaisälä frequency $\Omega_{\mathrm{max}} = (10.5 \pm 1)~\mathrm{rad~s^{-1}}$, and eventually the thermal expansion coefficient $\beta_T=\Omega_{\mathrm{max}}^2 /(\vec{g}\cdot\vec{\nabla} T)=(0.007 \pm 0.0013)~\mathrm{K^{-1}}$, as reported with the vertical dashed blue line in Fig. \ref{fig:SF1}(b).
All the results presented until here are coherent with those obtained in liquid binary or ternary mixtures for concentration fluctuations \cite{Croccolo_2019, garciafernandez_2019, ortizdezarate_2020}, thus confirming the possibility of extending those results to the study of temperature fluctuations in a supercritical fluid if the non equilibrium state is far enough from the critical point or a Widom line.

\bmhead{Case 2}

\begin{figure*} [ht!]
\centering
    \includegraphics[width=0.95\textwidth]{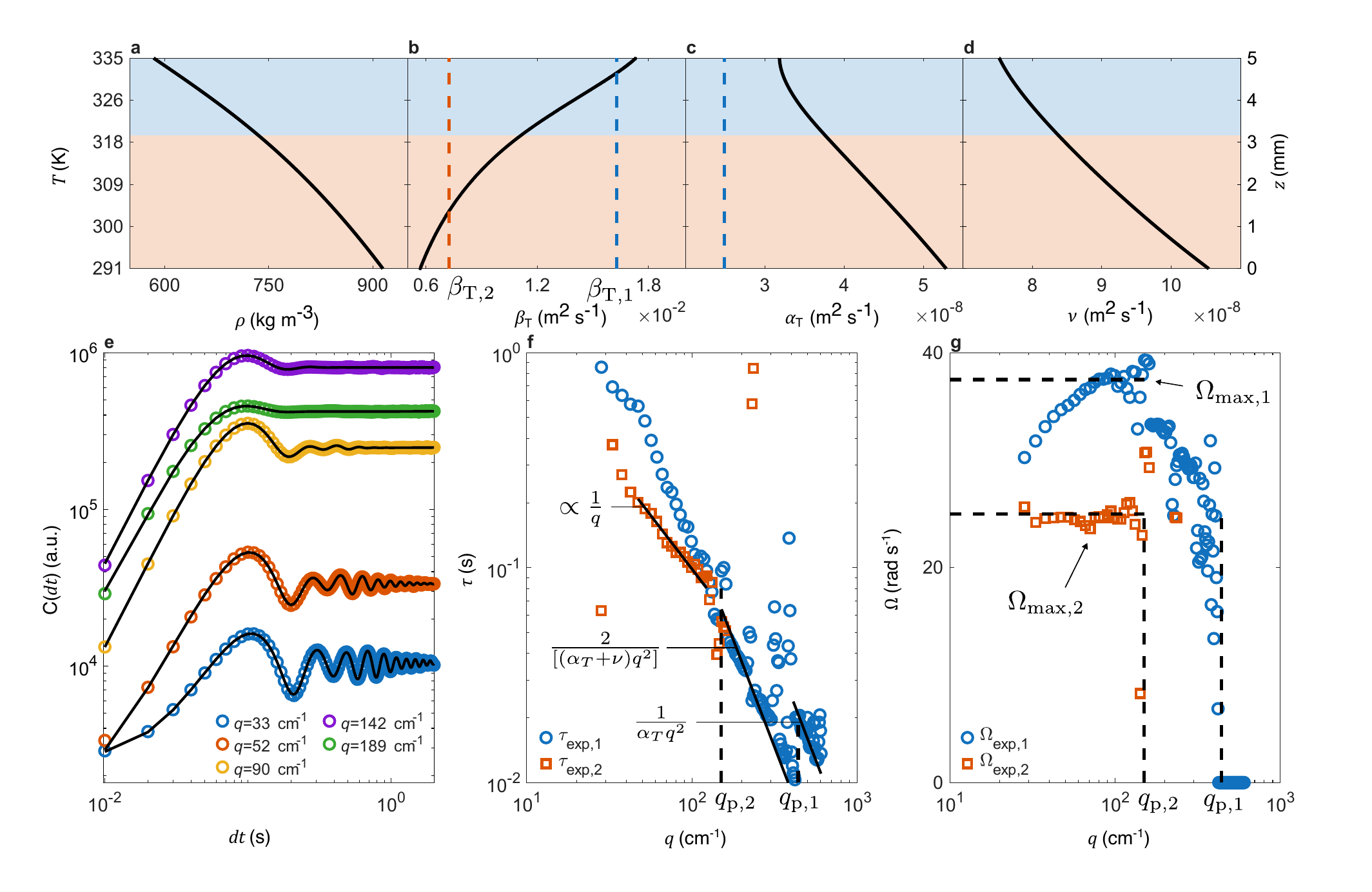}
    \caption{\textbf{Experimental results for case 2 under thermodynamic conditions crossing the Widom region.} Data at $p$ = 15 MPa, $T_{\mathrm{mean}}$ = 40 °C, $\Delta T$ = 44 K, green gradient in Fig.\ref{fig:densMap}.
        (a-d) Thermophysical properties of the system inside the cell: (a) density, (b) thermal expansion coefficient, (c) thermal diffusivity, and (d) kinematic viscosity. All the properties are plotted against temperature or height assuming a linear relationship between them. The temperature and height are on the vertical axis for all graphs. 
        The vertical dotted lines show the measured data from the fitting of the SFs, as explained in the text.
        (e) 1D-Structure functions for 5 wave numbers (33 (blue), 52 (orange), 90 (yellow), 142 (purple), and 189 (green) $\mathrm{cm^{-1}}$) and their fits (dark line). (f) Time decays $\tau(q)$ of the density fluctuations and (g) Brunt-Vaisälä frequencies $\Omega(q)$ as obtained by fitting the structure functions shown in (e) with the two order model of Equation (\ref{eq:ISF_2exp}).}
    \label{fig:SF2}
\end{figure*}

The second experiment explores thermodynamic conditions crossing the Widom region, where the thermophysical properties of the fluid are no longer linear with respect to temperature. This is achieved under the same pressure of the first experiment ($p$ = 15 MPa), but with a larger thermal gradient ($T_{\mathrm{mean}}$ = 40 °C, $\Delta T = 44~\mathrm{K}$, green gradient in Fig.~\ref{fig:densMap}). The resulting SFs plotted in Fig.~\ref{fig:SF2}(e) for different $q$ values, show a different behaviour that cannot be captured by the model structure function of Equation (\ref{eq:sf_with_1_oscillation}). 
Indeed, $\beta_T$, $\alpha_T$, and $\nu$ exhibit substantial relative variations $\Delta_{\mathrm{rel}}\beta_T = 114\%$, $\Delta_{\mathrm{rel}}\alpha_T =51\%$, $\Delta_{\mathrm{rel}}\nu =34\%$. Under these conditions, the assumption of a single effective base state $(T_0, p_0)$ is no longer valid across the entire layer. Following the discretize formulation of Equation (\ref{eq:ISFdisc}), the measured ISF can be approximated by two dominant contributions corresponding to two base states taken at different depths. It is now fitted with
\begin{equation}
\begin{split}
    f(q,dt)=&(1-a)\frac{\mathrm{cos}\big[\Omega_1(q)~dt+\phi_1\big]}{\mathrm{cos}(\phi_1)}~\mathrm{exp}\left[-\frac{dt}{\tau_1(q)}\right]\\
    &+a\frac{\mathrm{cos}\big[\Omega_2(q)~dt+\phi_2\big]}{\mathrm{cos}(\phi_2)}~\mathrm{exp}\left[-\frac{dt}{\tau_2(q)}\right].
    \label{eq:ISF_2exp}
\end{split}
\end{equation}
The main idea is that a continuous superposition of modes stemming from the infinitesimal layers within the sample needs to be taken into account, and the resulting ISF can be approximated by two main modes, similarly to what is done in the cumulant method \cite{pusey2015}, or in other, more refined methods \cite{Castellini_2023}. 
In this case, then, the fitting of the SFs provides two distinct time decays as well as two distinct oscillation frequencies, at least for a certain range of wave numbers. The resulting time decays $\tau_1(q)$ and $\tau_2(q)$ and frequencies $\Omega_1(q)$ and $\Omega_2(q)$ are presented in Fig. \ref{fig:SF2}(f) and (g), respectively.

Similarly to the first experiment, the time decay $\tau_1(q)$ plotted in blue in Fig. \ref{fig:SF2}(f) can be fitted for $q>q_{p,1}$ so to obtain a measurement of the fluid thermal diffusivity $\alpha_T=(2.7 \pm 0.3)\times 10^{-8}~\mathrm{m^2~ s^{-1}}$, a value reported with the vertical dashed blue line in Fig. \ref{fig:SF2}(c). For wave numbers $q<q_{p,1}$ the fitting provides an average of the thermal diffusivity and the viscosity, from which the kinematic viscosity can be derived: $\nu=(11 \pm 2)\times 10^{-8}~\mathrm{m^2~s^{-1}}$, as shown in Fig. \ref{fig:SF2}(d).

The time decay values $\tau_2(q)$, reported as orange squares in Fig. \ref{fig:SF2}(f) for values of $q<q_{p,2}$ correspond to a second oscillatory mode and are compatible with a sub-diffusive behaviour $\tau(q)\propto q^{-1}$ that cannot be explained according to existing models. Further analysis of this behaviour is outside the scope of this Letter.

The analysis of the frequencies reported in Fig. \ref{fig:SF2}(g) reveals maximum values of $\Omega_{\mathrm{1,max}}=(37.5 \pm 1)~\mathrm{rad.s^{-1}}$ and $\Omega_{\mathrm{2,max}}=(25 \pm 1)~\mathrm{rad.s^{-1}}$. Using the same approach as above, we obtain two different values of the thermal expansion coefficient $\beta_{T,1}=(0.0163 \pm 0.001)~\mathrm{K^{-1}}$ and $\beta_{T,2}=(0.0072 \pm 0.0006)~\mathrm{K^{-1}}$, as reported with vertical dashed lines in Fig. \ref{fig:SF2}(b).

The presence of two distinct propagating modes at different frequencies can be interpreted as a signature of fluid stratification in layers with different density gradients and, consequently, different $\beta_T$, which can be effectively measured using the described method. This result is particularly striking because the optical method provides integrated information across the fluid thickness. Starting from the experimental values of $\beta_T$, we define two distinct volumes by choosing the separation height as the one for which averaging the theoretical values of $\beta_T$ over two vertical layers separated at that height provides average $\beta_T$'s closest to the experimental values.

\bmhead{Case 3}

\begin{figure*} [ht!]
\centering
    \includegraphics[width=0.95\textwidth]{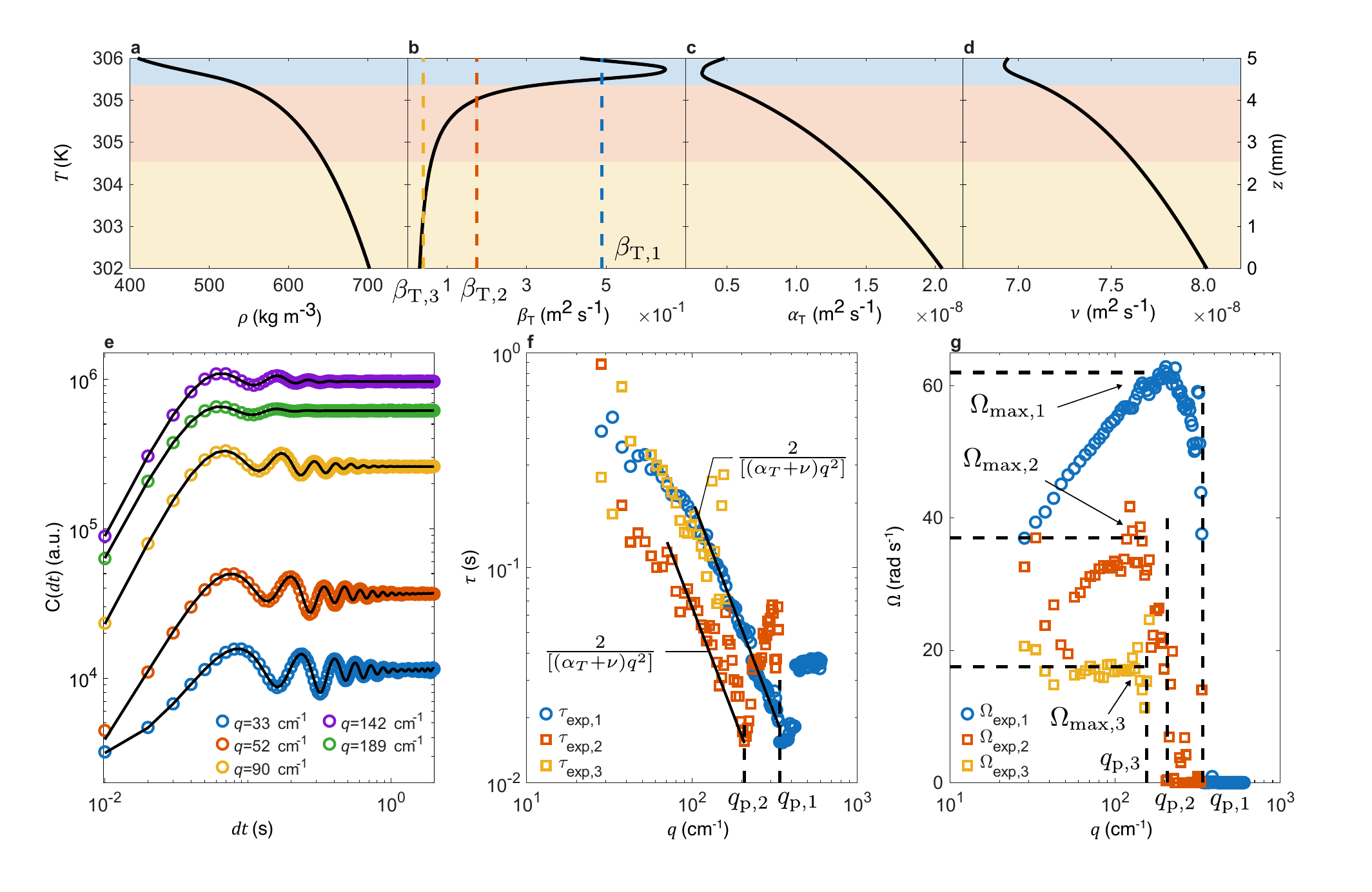}
    \caption{\textbf{Experimental results for case 3 under thermodynamic conditions crossing the Widom region, close to the critical point.} Data at $p$ = 7.7 MPa, $T_{\mathrm{mean}}$ = 31.1 °C, $\Delta T$ = 4 K, red gradient in Fig.\ref{fig:densMap}.
    (a-d) Thermophysical properties of the system inside the cell: (a) density, (b) thermal expansion coefficient, (c) thermal diffusivity, and (d) kinematic viscosity. All the properties are plotted against temperature or height supposing a linear relationship between them. The temperature and height are on the vertical axis for all graphs.
    (e) 1D-Structure functions for 5 wave numbers (33 (blue), 52 (orange), 90 (yellow), 142 (purple), and 189 (green) $\mathrm{cm^{-1}}$) and their fits (dark line). (f) Time decays $\tau$ of the density fluctuations and (g) Brunt-Väisälä frequencies $\Omega$ as obtained by fitting the structure functions shown in (e) with a three order model of Equation (\ref{eq:ISFdisc}).}
    \label{fig:SF3}
\end{figure*}

Finally, we report the results of a third experiment under thermodynamic conditions closer to the critical point, $p$ = 7.7 MPa, $T_{\mathrm{mean}}$ = 31.1 °C, $\Delta T$ = 4 K, red thermal gradient in Fig.\ref{fig:densMap}, thus crossing the isochore line, as visible in the inset. In such conditions, the density profile is drastically non-linear, and its gradient, or equivalently $\beta_T$, shows a pronounced peak across the Widom line, as visible in Fig. \ref{fig:SF3}(b). The resulting relative variations are: $\Delta_{\mathrm{rel}}\beta_T = 459\%$, $\Delta_{\mathrm{rel}}\alpha_T = 136\%$, $\Delta_{\mathrm{rel}}\nu =15\%$.

The resulting structure functions, reported in Fig. \ref{fig:SF3}(e), can be modelled neither by one propagating mode, as in Equation (\ref{eq:sf_with_1_oscillation}), nor by the sum of two propagating modes, as in Equation (\ref{eq:ISF_2exp}). Using the discrete procedure of Equation (\ref{eq:ISFdisc}), three layers are required to obtain reasonable results from the fitting procedure. In this case the fitting provides three time decays $\tau_1(q)$, $\tau_2(q)$ and $\tau_3(q)$ shown in Fig. \ref{fig:SF3}(f) and three oscillation frequencies $\Omega_1(q)$, $\Omega_2(q)$ and $\Omega_3(q)$ shown in Fig. \ref{fig:SF3}(g). Since fitting becomes rather complex with such a model, fitting can be performed only for $q<q_{p,1}$. As visible in the graph, the time decays $\tau_1(q)$ and $\tau_3(q)$ are essentially overlapping over a wide range of wave numbers. The fitting of their values again provides an average between the thermal diffusivity and the kinematic viscosity of the fluid.

The origin of such different behaviour can be better identified by analysing the oscillation frequencies shown in Fig. \ref{fig:SF3}(g). Three distinct oscillations can be clearly identified with three different maximum frequencies, $\Omega_{\mathrm{1,max}}=(62 \pm 2)~\mathrm{rad.s^{-1}}$, $\Omega_{\mathrm{2,max}}=(37 \pm 4)~\mathrm{rad.s^{-1}}$ and $\Omega_{\mathrm{3,max}}=(17.5 \pm 1)~\mathrm{rad.s^{-1}}$, corresponding to 
three values of $\beta_{T,1}=(0.490 \pm 0.03)~\mathrm{K^{-1}}$, $\beta_{T,2}=(0.174 \pm 0.04)~\mathrm{K^{-1}}$ and $\beta_{T,3}=(0.039 \pm 0.005)~\mathrm{K^{-1}}$, as reported with vertical dashed lines in Fig. \ref{fig:SF3}(b). These experimental measurements of the thermal expansion coefficients are compatible with literature values over the sublayers shown in Fig.\ref{fig:SF3}(b). This allows us to suggest the origin of the oscillations in different layers, that can be referred to as a quasi-liquid layer at the bottom of the sample cell for $\beta_{T,3}$, plus two intermediate layers where $\beta_T$ changes drastically, which is the signature of the transition towards a quasi-gaseous phase that is not present in the volume under the reported experimental conditions.

The latter interface layer does not show any singularity of its thermophysical properties, as it happens at the (nanometric) interface of a subcritical two-phase system. Instead, we can interpret it as a thick region where the thermophysical properties vary continuously, although strongly. This results in a vanishing but measurable surface tension between the two adjacent layers, an analysis that is beyond the scope of the current work.

\section*{Discussion}
These experiments reveal three main points. First, in supercritical \ce{CO2}, the interplay of thermal diffusion, viscosity, gravity and the presence of macroscopic gradients leads to depth-dependent dynamical behaviour in the fluctuation spectrum, which can be interpreted as a spontaneous stratification into quasi-phases. Indeed, multiple relaxation times and frequencies emerge from the fitting of the structure functions, which correlate quantitatively with depth-dependent thermophysical properties. This effect is more pronounced when crossing the Widom region, i.e., where the density gradients are larger. Second, in a single non-equilibrium experiment we can test a continuum of points in the ($p$, $T$) phase space of the fluid system, thus detecting different behaviours of the supercritical fluid. Therefore, third point, we are able to obtain quantitative information about the fluid thermophysical properties stemming from different depths of the volume and separating them in the time (frequency) domain.

These results provide additional insight into the behaviour of fluids in the supercritical and Widom regions. Whereas two-state and Kirkwood--Buff-based frameworks \cite{Kirkwood1951,Smith2006,Ploetz2019} provide equilibrium structural interpretations of the Widom region, our measurements probe hydrodynamic fluctuations under non-equilibrium temperature gradients and do not directly access microscopic structural order. The present results are therefore interpreted within the framework of fluctuating hydrodynamics. The possible connection between the observed dynamical stratification and underlying microscopic structural heterogeneity remains an open question.

According to our results, the Widom region of supercritical \ce{CO2} cannot be understood as a homogeneous state. Instead, the fluid subjected to a macroscopic temperature gradient and to the action of buoyancy exhibits stratified structures whose properties vary dramatically. These findings indicate that the classic equilibrium thermodynamic approach alone can be insufficient to describe the dynamical structure of a supercritical fluid under non-equilibrium gradients. 

The observation of multiple Brunt–Väisälä frequencies highlights a coupling between thermal and viscous modes that has no analogue in equilibrium fluids. These results suggest a new framework in which the supercritical state is viewed as a set of stratified hydrodynamic layers exhibiting various dynamics. Such a perspective has direct implications for Carbon Capture, Utilisation and Storage (CCUS) strategies \cite{imre_anomalous_2015}, energy systems exploiting supercritical \ce{CO2} \cite{ren_boundary-layer_2019,yu2021}, and planetary sciences where non-equilibrium gradients are ubiquitous.
It would be intriguing to perform a similar experiment under reduced-gravity conditions, to check whether the fluid stratification would persist. That would confirm that the stratification is dominated by temperature rather than gravity. Its stability could be influenced by the presence of giant fluctuations of the temperature, and thus the density that could diverge, not being damped by the buoyancy force \cite{Vailati2020}. Recent advancements in the modelling of phase changes through fluctuating hydrodynamics \cite{Gallo_2023} show the potential of our approach also below the critical point.

\section*{Methods}\label{sec5}
\bmhead{Experimental setup}
Experiments are conducted in a high-pressure optical cell filled with \ce{CO2} at supercritical conditions (temperature in the range 291.15–335.15 K, pressures above 7.4 MPa). A vertical temperature gradient is applied across the fluid layer using Peltier elements controlled with PID regulation. Density fluctuations are imaged using dynamic shadowgraphy with a superluminescent diode illumination source and a s-CMOS camera. Image sequences are analysed using differential dynamic analysis (DDA) \cite{Croccolo2012}, yielding structure functions over spatial wave numbers and time delays. Fits of intermediate scattering functions extracted characteristic decay times and Brunt–Väisälä frequencies.

\bmhead{Pressure control and monitoring}
The HP cell is pressurized with \ce{CO2} with the help of a high-pressure syringe pump (500D, Teledyne ISCO), which allows for monitoring the pressure inside the sample during the procedure. When the thermodynamic base state $(T_0,p_0)$ is reached and stabilized, the thermal gradient is applied to the fluid layer. Once the stationary state is reached, we isolate the cell from the Teledyne ISCO pump by closing a valve. The volume is now only connected to the pressure sensor (33X Series, from Keller) that reads and records the pressure via a K-114 convertor interface. During an experiment, the recorded pressure remains constant, with fluctuations of less than 0.02 MPa.

\bmhead{Thermophysical properties}
To compare our experimental values with theoretical ones, we refer to the NIST Chemistry WebBook which provides the density $\rho$ with a precision of 0.05\% \cite{Span1996}, the specific heat mass $c_p$ (1.5\%) \cite{Span1996}, the thermal conductivity $k$ (2\%) \cite{Huber2016}, and the dynamic viscosity $\mu$ \cite{Laesecke2017}. These literature values allow us to compute the thermal expansion coefficient $\beta_T = \frac{1}{\rho}\frac{\partial\rho}{\partial T}$ and the thermal diffusivity $\alpha_T = \frac{k}{\rho c_p}$ as a function of temperature and pressure, i.e. according to the position within the sample making the assumption of a linear temperature gradient.

\bmhead{Fitting procedure and robustness}

For each case, the temporal structure factors $S(q,dt)$ were fitted using the discrete-mode model of Equation ~\ref{eq:ISFdisc}), testing various models with up to four oscillatory damped decaying modes. The fitting sequence was initialized at the anchor wave number $q=162~\mathrm{cm^{-1}}$, corresponding to the first maximum of the shadowgraph transfer function, where the signal-to-noise ratio is the largest, see Fig.\ref{fig:images} (d). At this wave number, a strong robustness analysis, described below, was performed. The fitting was then propagated iteratively toward smaller and larger $q$ values by using the results of the fit at the previous wave number as the initial guess. This approach is justified by the smooth dependence of the fitted parameters with $q$, particularly the decay times $\tau_i$ and oscillation frequencies $\Omega_i$.

At each iteration, the fit was performed using the weighting function $w(dt)=\exp(-dt)$, which emphasizes short-time dynamics where damped oscillatory features are more pronounced. The fitting temporal window was limited to $dt_{\max}=2~\text{s}$, as the theoretically expected $\tau$ values are smaller over the explored $q$-range. The cut-off wave number $q_p$ was estimated over the temperature range using NIST data, and its maximum value within the bulk was identified. This allows preselecting whether using an oscillatory($q<q_p$) or non-oscillatory ($q>q_p$) fitting model.

To evaluate the quality of the fits and select the best model, we used a weighted relative residual sum of squares (relRSS)
\begin{equation}
\mathrm{relRSS} = \frac{\sum_{i=1}^{n} w_i \left( y_i^{\mathrm{exp}} - y_i^{\mathrm{fit}} \right)^2}{\sum_{i=1}^{n} w_i \, y_i^{\mathrm{exp}}\,^2}.
\end{equation}
At each wave number, the selected model was the lowest-order model whose relRSS fell below a threshold of $10^{-4}$ or $10^{-5}$, depending on the case studied. We also ensure that the modes are statistically independent by not retaining those with very close $\tau_i$ and $\Omega_i$, or with an amplitude $a_i < 0.01$ that does not represent a meaningful contribution. 

The robustness of the procedure is first tested at the anchor wave number $q/q_{\min}=34$. For each model order $N=1,\dots,4$, the fit is repeated using three weighting functions ($1$, $1/dt$, and $\exp(-dt)$), three fitting windows ($dt_{\max}=2,5, \text{or}~10~\text{s}$), and 100 random initial conditions, yielding 900 fits per model. For each model, we compute the success rate and the mean and median of relRSS. The success rate is here defined as the ratio between the fits with a relRSS lower than the threshold and the total number of statistical independent fits. We further quantified robustness using the mean and standard deviation of the dominant mode parameters $\tau_1$ and $\Omega_1$ across the selected fits. The results for all three cases are summarized in Table~\ref{tab:anchor_robustness_all_cases}. The anchor model is selected as the lowest-order model that combines a high success rate, low relRSS, and minimal parameter dispersion.

Robustness across the full $q$-range is ensured by enforcing continuity of $\tau_i(q)$ and $\Omega_i(q)$, using the solution at the previous $q$ as initialization. For higher-order models, we ensure robustness by sampling multiple randomized initial conditions. 

\begin{table*}[t]
\centering
\caption{Anchor robustness analysis for the three cases. For each model order, we report the success rate, the median and mean relative residual sum of squares, and the mean and standard deviation of the dominant mode parameters $\tau_1$ and $\Omega_1$. For each case, the selected model is shown in bold. }
\label{tab:anchor_robustness_all_cases}
\begin{tabular}{c c c c c c c c}
\hline
Modes & Success rate & Med. relRSS & Mean relRSS & $\overline{\tau_1}$ (s) & $\sigma_{\tau_1}$ (s) & $\overline{\Omega_1}$ (rad s$^{-1}$) & $\sigma_{\Omega_1}$ \\
\hline
\hline
\multicolumn{8}{c}{\textbf{Case 1}} \\
\hline
\textbf{1} & \textbf{63\%} & \textbf{1.2e-5} & \textbf{8.5e-3} & \textbf{0.057} & \textbf{0.0023} & \textbf{8.9} & \textbf{0.41} \\
2 & 81\% & 5.8e-7 & 2.6e-3 & 0.071 & 0.069 & 14 & 4 \\
3 & 91\% & 5.1e-7 & 1.1e-3 & 0.074 & 0.12 & 25 & 11 \\
4 & 92\% & 3.3e-7 & 6.7e-4 & 0.54 & 2.5 & 31 & 12 \\
\hline
\multicolumn{8}{c}{\textbf{Case 2}} \\
\hline
1 & 0\% & 1.5e-5 & 1.5e-3 & $\mathrm{N/A}$ & $\mathrm{N/A}$ & $\mathrm{N/A}$ & $\mathrm{N/A}$ \\
\textbf{2} & \textbf{69\%} & \textbf{5.7e-6} & \textbf{2.2e-4} & \textbf{0.067} & \textbf{0.016} & \textbf{39} & \textbf{8.6} \\
3 & 72\% & 2.8e-6 & 5.9e-4 & 0.061 & 0.014 & 38 & 5.9 \\
4 & 74\% & 1.6e-7 & 3.7e-4 & 0.25 & 1.5 & 41 & 6.6 \\
\hline
\multicolumn{8}{c}{\textbf{Case 3}} \\
\hline
1 & 0\% & 5.8e-4 & 1.7e-3 & $\mathrm{N/A}$ & $\mathrm{N/A}$ & $\mathrm{N/A}$ & $\mathrm{N/A}$ \\
2 & 43\% & 1.0e-5 & 2.9e-4 & 0.074 & 0.0011 & 60 & 0.35 \\
\textbf{3} & \textbf{88\%} & \textbf{2.5e-6} & \textbf{4.2e-4} & \textbf{0.08} & \textbf{0.0084} & \textbf{59} & \textbf{2.3} \\
4 & 85\% & 8.4e-7 & 2.8e-4 & 0.079 & 0.025 & 60 & 4.1 \\
\hline
\end{tabular}
\end{table*}

\bmhead{Data Availability}
The processed data for the plots generated in this study have been deposited in the figshare database under accession link \hyperlink{https://doi.org/10.6084/m9.figshare.32408229}{https://doi.org/10.6084/m9.figshare.32408229}. They are also available in the Source Data File. Additional raw image data are available from the corresponding authors upon reasonable request owing to the large storage requirements associated with these files.


\section*{Acknowledgements}
This research was conducted within the E2S UPPA Hub Newpores and the Industrial Chair CO2ES, supported by the Investissements d’Avenir French program managed by ANR (No. ANR16IDEX0002). Funding from CNES through the CNRS GdR 2799 MFA is also acknowledged. We thank Guillaume Galliero and Alberto Vailati for fruitful discussions.

\section*{Author contributions}
C.G. conceived the project and P.F. and F.C. supervised it. P.F., E.L. and C.G. designed and executed the experiments. P.F., H.I., C.G. and F.C. analysed and discuss the data. P.F. and F.C. wrote the paper with insights from H.I. and C.G.. P.F., E.L., H.I., C.G. and F.C. reviewed the paper. F.C. funded and managed the project.

\section*{Competing interest}
The authors declare no competing interests.

\end{document}